\documentclass[12pt]{article}
\begin{document}

\newenvironment{describe}{\begin{list}{}{\setlength\leftmargin{80pt}}\setlength\labelsep{10pt}\setlength\labelwidth{70pt}}{\end{list}}

\newenvironment{flag}{\begin{list}{\makebox[20pt]{\hss$\circ$\enspace}}
                                  {\labelwidth 20pt}}{\end{list}}

%% js \newtheorem{proposition}{Proposition}

%% js \newenvironment{proof}
     %% js {\begin{trivlist}\item[]{\bf Proof. }}%
     %% js {\\* \hspace*{\fill} $\Box$\end{trivlist}}

\newenvironment{numberedlist}
{\begin{list}{\makebox[20pt]{\hss(\arabic{itemno})\enspace}}
             {\usecounter{itemno}\labelwidth 20pt}}{\end{list}}

\newenvironment{alphabetlist}
{\begin{list}{\makebox[20pt]{\hss(\alph{itemno1})\enspace}}
             {\usecounter{itemno1}\labelwidth 20pt}}{\end{list}}

\newenvironment{romanlist}
{\begin{list}{\makebox[20pt]{\hss(\roman{itemno2})\enspace}}
             {\usecounter{itemno2}\labelwidth 20pt}}{\end{list}}

\newcounter{itemno}

\newcounter{itemno1}

\newcounter{itemno2}
\newcounter{lemma}
\newcounter{exno}

\newcounter{defno}

%\newcounter{exno}[section]

%\newcounter{defno}[section]

%\newtheorem{defn}{Definition}[section]

%\newtheorem{ex}[defn]{Example}

%% js \newtheorem{lemma}{Lemma}

%% js \newtheorem{theorem}[lemma]{Theorem}

\newenvironment{defn}{\refstepcounter{defno}\medskip \noindent {\bf
Definition \thedefno.\ }}{\medskip}

\newenvironment{ex}{\refstepcounter{exno}\medskip \noindent {\bf
Example \theexno.\ }}{\medskip}

\newenvironment{millerexample}{
 \begingroup \begin{tabbing} \hspace{2em}\= \hspace{5em}\= \hspace{5em}\=
\hspace{5em}\= \kill}{
 \end{tabbing}\endgroup}

\newenvironment{wideexample}{
 \begingroup \begin{tabbing} \hspace{2em}\= \hspace{10em}\= \hspace{10em}\=
\hspace{10em}\= \kill}{
 \end{tabbing}\endgroup}

\newcommand{\sep}{\;\vert\;}

\newcommand{\ra}{\rightarrow}
\newcommand{\app}{\ }
\newcommand{\appt}{\ }
\newcommand{\tup}[1]{\langle\nobreak#1\nobreak\rangle}

\newcommand{\hu}{{\cal H}^+}
\newcommand{\Free}{{\cal F}}
\newcommand{\oprove}{\vdash\kern-.6em\lower.7ex\hbox{$\scriptstyle O$}\,}
\newcommand{\true}{\top}

\newcommand{\Dscr}{{\cal D}}
\newcommand{\Pscr}{{\cal P}}
\newcommand{\Gscr}{{\cal G}}
\newcommand{\Fscr}{{\cal F}}
\newcommand{\Vscr}{{\cal V}}
\newcommand{\Uscr}{{\cal U}}
\newcommand{\pderivation}{{\cal P}\kern -.1em\hbox{\rm -derivation}}
\newcommand{\pderivationl}{{\cal P}\kern -.1em\hbox{\em -derivation}}
\newcommand{\pderivable}{{\cal P}\kern -.1em\hbox{\rm -derivable}}
\newcommand{\pderivablel}{{\cal P}\kern -.1em\hbox{\em -derivable}}
\newcommand{\pderivations}{{\cal P}\kern -.1em\hbox{\rm -derivations}}
\newcommand{\pderivability}{{\cal P}\kern -.1em\hbox{\rm -derivability}}
\newcommand{\eqm}[1]{=_{\scriptscriptstyle #1}}
\newcommand\subsl{\preceq}
\newcommand{\fnrestr}{\uparrow}

\newcommand{\match}{{\rm MATCH}}
\newcommand{\triv}{{\rm TRIV}}
\newcommand{\imit}{{\rm IMIT}}
\newcommand{\proj}{{\rm PROJ}}
\newcommand{\simpl}{{\rm SIMPL}}
\newcommand{\failed}{{\bf F}}

\newcommand{\Dsiginst}[1]{{[#1]_\Sigma}}
\newcommand{\Psiginst}[1]{{[#1]_\Sigma}}
\newcommand{\lnorm}{{\lambda}norm}
\newcommand{\seq}[2]{#1 \supset #2}
\newcommand{\dseq}[2]{#1_1,\ldots,#1_{#2}}

\newcommand{\all}{\forall}
\newcommand{\some}{\exists}
\newcommand{\lambdax}[1]{\lambda #1\,}
\newcommand{\somex}[1]{\some#1\,}
\newcommand\allx[1]{\all#1\,}

\newcommand{\subs}[3]{[#1/#2]#3}
\newcommand{\rep}[3]{S^{#2}_{#1}{#3}}
\newcommand{\ie}{{\em i.e.}}
\newcommand{\eg}{{\em e.g.}}

% These are the annotations used with inference figures
\newcommand{\lbotr}{$\bot$-R}
\newcommand{\ldbotr}{\bot\mbox{\rm -R}}
\newcommand{\landl}{$\land$-L}
\newcommand{\ldandl}{\land\mbox{\rm -L}}
\newcommand{\landr}{$\land$-R}
\newcommand{\ldandr}{\land\mbox{\rm -R}}
\newcommand{\lorl}{$\lor$-L}
\newcommand{\ldorl}{\lor\mbox{\rm -L}}
\newcommand{\lorr}{$\lor$-R}
\newcommand{\ldorr}{\lor\mbox{\rm -R}}
\newcommand{\limpl}{$\supset$-L}
\newcommand{\ldimpl}{\supset\mbox{\rm -L}}
\newcommand{\limpr}{$\supset$-R}
\newcommand{\ldimpr}{\supset\mbox{\rm -R}}
\newcommand{\lnegl}{$\neg$-L}
\newcommand{\ldnegl}{\neg\mbox{\rm -L}}
\newcommand{\ldnegr}{\neg\mbox{\rm -R}}
\newcommand{\lalll}{$\forall$-L}
\newcommand{\ldalll}{\forall\mbox{\rm -L}}
\newcommand{\lallr}{$\forall$-R}
\newcommand{\ldallr}{\forall\mbox{\rm -R}}
\newcommand{\lsomel}{$\exists$-L}
\newcommand{\ldsomel}{\exists\mbox{\rm -L}}
\newcommand{\lsomer}{$\exists$-R}
\newcommand{\ldsomer}{\exists\mbox{\rm -R}}
\newcommand{\ldlamlr}{\lambda}
\newcommand{\sequent}[2]{\hbox{{$#1\ \longrightarrow\ #2$}}}
\newcommand{\prog}[2]{\hbox{{$#1\ \supset\ #2$}}}
\newcommand{\run}{\Gamma}

\newcommand{\Ibf}{{\bf I}}
\newcommand{\Cbf}{{\bf C}} 
\newcommand{\Cbfpr}{{\bf C'}}

\newcommand{\cprove}{\vdash_C}
\newcommand{\iprove}{\vdash_I}

\newsavebox{\lpartfig}
\newsavebox{\rpartfig}

% From the hohh section

\newenvironment{exmple}{
 \begingroup \begin{tabbing} \hspace{2em}\= \hspace{3em}\= \hspace{3em}\=
\hspace{3em}\= \hspace{3em}\= \hspace{3em}\= \kill}{
 \end{tabbing}\endgroup}
\newenvironment{example2}{
 \begingroup \begin{tabbing} \hspace{8em}\= \hspace{2em}\= \hspace{2em}\=
\hspace{10em}\= \hspace{2em}\= \hspace{2em}\= \hspace{2em}\= \kill}{
 \end{tabbing}\endgroup}

\newenvironment{example}{
\begingroup  \begin{tabbing} \hspace{2em}\= \hspace{3em}\= \hspace{3em}\=
\hspace{3em}\= \hspace{3em}\= \hspace{3em}\= \hspace{3em}\= \hspace{3em}\= 
\hspace{3em}\= \hspace{3em}\= \hspace{3em}\= \hspace{3em}\= \kill}{
 \end{tabbing} \endgroup }

\newcommand{\sand}{sand} % choice disjunction
\newcommand{\pand}{pand} % choice disjunction
\newcommand{\cor}{cor} % choice disjunction

\newcommand{\lb}{\langle}
\newcommand{\rb}{\rangle}
\newcommand{\pr}{prov}
\newcommand{\prG}{intp}
\newcommand{\prSG}{intp_E}
\newcommand{\intp}{intp_o}
\newcommand{\prove}{exec} % choice conjunction
\newcommand{\np}{invalid} % choice conjunction
\newcommand{\Ra}{\supset}  
\newcommand{\add}{\oplus} % choice disjunction
\newcommand{\adc}{\&} % choice conjunction
\newcommand{\Cscr}{{\cal C}}
\newcommand{\seqweb}{SProlog}
\newcommand{\sprog}{{SProlog}}

\newtheorem{theorem}[lemma]{Theorem}

\newtheorem{proposition}[lemma]{Proposition}

\newtheorem{corollary}[lemma]{Corollary}
\newenvironment{proof}
     {\begin{trivlist}\item[]{\it Proof. }}%
     {\\* \hspace*{\fill} \end{trivlist}}

\newcommand{\seqand}{\prec}
\newcommand{\seqor}{\cup}
\newcommand{\seqandq}[2]{\prec_{#1}^{#2}}
\newcommand{\parandq}[2]{\land_{#1}^{#2}}
\newcommand{\exq}[2]{\exists_{#1}^{#2}}
\newcommand{\ext}{intp_G}

\newcommand{\muprolog}{Prolog$_{\add}$}
\newcommand{\cdc}{choice-disjunctive clauses}
\renewcommand{\pr}{pv}
\newcommand{\pra}{pv_D^\Uparrow} % choice conjunction
\newcommand{\prs}{pv_D^\Downarrow} 
\newcommand{\prg}{pv_G} 

\newcommand{\pe}{ex}
\newcommand{\pea}{ex_D^\Uparrow} % choice conjunction
\newcommand{\pes}{ex_D^\Downarrow}
\newcommand{\peg}{ex_G} 

%\begin{document}

\begin{center}
{\Large {\bf Interactive Logic Programming via
Choice-Disjunctive Clauses }}
\\[20pt] 
{\bf Keehang Kwon }\\
Dept. of Computer  Engineering, DongA University \\
Busan 604-714, Korea\\
%051-200-7784 \\
 khkwon@dau.ac.kr\\
\end{center}

\noindent {\bf Abstract}: 
Adding interaction to  logic programming  is an essential task. 
 Expressive logics such as linear logic provide a theoretical basis for such a
 mechanism.
 Unfortunately, none of the existing linear logic languages can model interactions with the user. This is because
they uses provability as the sole basis for computation.

   We propose to use the game semantics instead of provability
   as the basis for computation  to allow for more active participation from the user. We illustrate our idea
via \muprolog, an extension of Prolog with choice-disjunctive clauses.
%\end{summary}

{\bf keywords:} interaction, logic programming, linear logic, computability logic.

\section{Introduction}\label{sec:intro}

Representing interactive objects (lottery tickets, vending machines)  in  logic and logic programming requires interactive knowledgebases or
interactive clauses. An interactive 
knowledgebase must be able to allow
the user to select one among many alternatives. Expressive logics such as linear logic provide a theoretical basis for such a
 mechanism.

Unfortunately, none of the existing linear logic languages can model decision steps from the user.
 This deficiency is an outcome of
using  provability as the sole basis for executing logic programs.
In the operational semantics based on  provability such as  uniform provability 
\cite{HM94,Mil89jlp,MNPS91}, 
solving a goal $G$ from the additive-disjunctive clause $D_0 \add D_1$ 
 simply {\it terminates} with a success if $G$ is solvable from
 both $D_0$ and $D_1$.
This semantics, $\pr$, is shown below:

\[ \pr(D_0\add D_1, G)\ if\ \pr(D_0, G)\ and\ \pr(D_1, G) \] 

\noindent
This is  unsatisfactory, as  the action of choosing either $D_0$ 
or $D_1$ by the
user --  the declarative reading  of $\add$ -- is not present in this operational semantics.

 Our approach in this paper involves a change of the  operational
semantics to allow for more active participation from the user. This is inspired by the
game semantics of Japaridze \cite{Jap03}.
Solving a goal $G$ from the choice-disjunctive clause $D_0 \add D_1$  now  has the
following operational semantics:

\[ \pe(D_0\add D_1, G)\ if\ read(k)\ and\ \pe(D_i,G)\ and\  
 \pr(D_j,G) \] 

\noindent where  $i\ (=\ 0\ {\rm or}\ 1))$ is chosen by the user (and stored in $k$)
 and $j$ is $(i+1)\ mod\ 2$.
In the above semantics, the system  requests the user to choose $i$ and then proceeds
with solving both the chosen goal, $G_i$,  and the unchosen goal, $G_j$.
Both executions must succeed.
It is worth noting that
 solving  $G_j$ must proceed using $\pr$ rather than $\pe$
to disallow further interactions  with the user.
It can be easily seen that our new semantics has the advantage over the old semantics:
the former respects the declarative reading of $\adc$
without losing completeness.

As an   
illustration of this approach, let us consider a  BMW car dealer web page
where you can get the information for BMW  models you choose.  For an engine, you can have a gasoline model or a diesel one.
 For a doortype, you can 
have a 2door or a 4door.
This is provided by the following definition:

\begin{exmple}
$! 2door \add ! 4door.$\\
$! diesel \add ! gas.$\\
$!\ bmw(120d)$ ${\rm :-}$ \> \hspace{6em}   $2door \otimes diesel$.\\ 
$!\ bmw(120)$ ${\rm :-}$ \> \hspace{6em}      $2door \otimes gas$.\\ 
$!\ bmw(320d)$ ${\rm :-}$ \> \hspace{6em}    $4door \otimes diesel$.\\ 
$!\ bmw(320)$ ${\rm :-}$ \> \hspace{6em}    $4door \otimes gas$.\\ 

\end{exmple}
\noindent Here, ${\rm :-}$ represents reverse implication.
The  definition above consists of  reusable resources, denoted by $!$.
 As a particular example, consider a goal task $\some x\ bmw(x)$. This goal would simply terminate
with no interactions from the user  in the context of \cite{HM94} as this goal is  solvable.  However, in our context,
  execution proceeds as follows: the system 
 requests the user to select a particular engine model and a doortype. 
After they -- 
 say, $2door, diesel$ -- is selected, 
  execution eventually terminates with $x = 120d$.

     As seen from the example above, \cdc\ can be used to model 
interactive decision tasks. 

   To present our idea as simple as possible, this paper focuses on muprolog,
 which is a variant of a  subset of Lolli\cite{HM94}. 
The former can be obtained
from the latter by (a) disallowing linear context and $\adc$ in the clauses,  (b) allowing only
          $\otimes$   in goal formulas, and (c) allowing $\add$ in
          the clauses. \muprolog\ can also be seen
as an extension of Prolog with \cdc, as
$\otimes$ in \muprolog\ corresponds to $\land$ of Prolog.

In this paper we present the syntax and semantics of this extended language, 
show some examples of its use. 
The remainder of this paper is structured as follows. We describe  \muprolog\
based on a first-order  clauses  in
the next section and Section 3. In Section \ref{sec:modules}, we
present some examples of  \muprolog.
Section~\ref{sec:conc} concludes the paper.

\section{\muprolog\ and Its Proof Procedure}\label{sec:logic1}

The extended language is a version of Horn clauses
 with \cdc. It is described
by $G$-, $C$- and $D$-formulas given by the syntax rules below:
\begin{exmple}
\>$G ::=$ \>  $A \sep   G \otimes  G \sep    \some x\ G $ \\   \\
\>$C ::=$ \>  $A  \sep G \supset A\ \sep \all x\ C $\\
\>$D ::=$ \>  $!C  \sep D \add D $\\
\end{exmple}
\noindent
In the rules above, $A$  represents an atomic formula.
A $C$-formula is called a Horn clause and a $D$-formula  is called a choice-disjunctive
 clause. 

In the transition system to be considered, $G$-formulas will function as 
queries and a list of $D$-formulas will constitute  a program. 
 We will  present a proof procedure  for this language  as sequent system. 
The rules for proving queries in our language are based on
two different phases. The first phase is that of processing \cdc, while
the second phase is that of  proving traditional  Prolog based on
uniform provability \cite{HM94,MNPS91}. Note that choice-disjunctive
clauses are processed first via $\pra$.
  Then,  execution in the second phase
proceeds just like traditional logic programming. To be
specific, execution in the second phase alternates between 
two subphases: the goal reduction subphase via $\prg$ (one  without a distinguished clause)
and the backchaining subphase via $\prs$ (one with a distinguished clause).
Below in the notation $\prs(D,
\Pscr,G)$,  the $D$ formula is a distinguished formula
(marked for backchaining). The symbol $::$ is a list constructor.

\begin{defn}\label{def:semantics}
Let $G$ be a goal and let $\Delta$ be a list of $D$-formulas
 and
let $\Pscr$ be a set of Horn clauses.
Then the task of proving $G$ from $\Delta$ -- $\pr(\Delta,G)$ -- is defined as follows:

\begin{numberedlist}

\item $\pr(\Delta,G)$ if $\pra(nil,\Delta,G)$.

\item    $\pra(\Pscr,!C::\Delta,G)$ if 
 $\pra(\{ !C \} \cup \Pscr,\Delta, G)$.
 
 \item    $\pra(\Pscr, (D_0 \add D_1)::\Delta,G)$ if 
 $\pra(\Pscr,D_0::\Delta , G)$  and $\pra(\Pscr,D_1::\Delta , G)$.

 \item    $\pra(\Pscr,nil,G)$ if 
 $\prg(\Pscr, G)$. \% switch to goal reduction mode

\item  $\prs(A,\Pscr,A)$. \% This is a success.

\item    $\prs((G_0\supset A),\Pscr,A)$ if 
 $\prg(\Pscr, G_0)$.

\item    $\prs(\all x D,\Pscr,A)$ if   $\prs([t/x]D,
\Pscr, A)$.

\item    $\prg(\Pscr,A)$ if   $D \in \Pscr$ and $\prs(D,\Pscr, A)$.
\% switch to backchaining mode

\item $\prg(\Pscr, G_0 \otimes G_1)$  if $\prg(\Pscr, G_0)$ and 
$\prg(\Pscr, G_1)$.

 \item $\prg(\Pscr, \some x G_0)$  if $\prg(\Pscr, [t/x]G_0)$.

\end{numberedlist}
\end{defn}
\noindent  
The notion of proof defined above 
is intuitive enough. The following theorem -- whose proof is rather obvious from the discussion in 
 \cite{HM94} and from the completeness of the focused proof system --
 shows the connection to linear logic.

\begin{theorem}
Let $\Delta$ be a program and $G$ be a goal in \muprolog. The procedure
 $\pr(\Delta,G)$ 
is a success if and only if $G$ follows from !$\Delta$ in   intuitionistic  linear logic.
\end{theorem}

\section{An Execution Model for  \muprolog}\label{sec:logic}

We now present an execution model for \muprolog. This execution model
is identical to the proof procedure in the previous section, except
for the way  \cdc\ are handled (Rule 3 below).

In the transition system to be considered, $G$-formulas will function as 
queries and a list of $D$-formulas will constitute  a program. 
 We will  present an operational semantics  for this language  as before. 

\begin{defn}\label{def:semantics}
Let $G$ be a goal and let $\Delta$ be a list of $D$-formulas
 and
let $\Pscr$ be a set of Horn clauses.
Then the task of executing $G$ from $\Delta$ -- $\pe(\Delta ,G)$ -- is defined as follows:

\begin{numberedlist}

\item $\pe(\Delta,G)$ if $\pea(nil,\Delta,G)$.

\item    $\pea(\Pscr,!C::\Delta,G)$ if 
 $\pea(\{ !C \} \cup \Pscr,\Delta, G)$.
 
 \item    $\pea(\Pscr, (D_0 \add D_1)::\Delta,G)$ if $read(i)$ and
 $\pea(\Pscr,D_i::\Delta, G)$  and $\pra(\Pscr,D_j::\Delta, G)$ where  $i\ (=\ 0\ {\rm or}\ 1))$ is chosen by the user and $j$ is $(i+1)\ mod\ 2$.

 \item    $\pea(\Pscr,nil,G)$ if 
 $\peg(\Pscr, G)$. \% switch to traditional logic programming

\item  $\pes(A,\Pscr,A)$. \% This is a success.

\item    $\pes((G_0\supset A),\Pscr,A)$ if 
 $\peg(\Pscr, G_0)$.

\item    $\pes(\all x D,\Pscr,A)$ if   $\pes([t/x]D,
\Pscr, A)$.

\item    $\peg(\Pscr,A)$ if   $D \in \Pscr$ and $\pes(D,\Pscr, A)$.

\item $\peg(\Pscr, G_0 \otimes G_1)$  if $\peg(\Pscr, G_0)$ and 
$\peg(\Pscr, G_1)$. 

 \item $\peg(\Pscr, \some x G_0)$  if $\peg(\Pscr, [t/x]G_0)$.

\end{numberedlist}
\end{defn}
\noindent  
In the above rules, the symbol $\add$  provides 
choice operations.

The following theorem -- whose proof is easily obtained from the fact that the modified rule does not affect the soundness and completeness  and can be
shown using an induction on the length of derivations --
 shows the connection
between our  operational semantics  and linear logic.

\begin{theorem}
Let $\Delta$ be a program and $G$ be a goal in \muprolog. Executing $\lb\Delta,G\rb$ -- $\pe(\Delta,G)$ -- terminates
with a success if and only if $G$ follows from $! \delta$ in  
\end{theorem}

\section{Examples }\label{sec:modules}

As an  example, let us consider the following interactive database which contains tuition
information for some university. The following tuition  charges are in effect for 
this year: \$40K for medical students, \$30K for engineering and
 \$20K for economics.

\begin{exmple}
$! med \add ! eng \add ! eco$. \\
$! tuition(40K) :- med$.\\
$! tuition(30K) :- eng$.\\
$! tuition(20K) :- eco$.\\
\end{exmple}
\noindent Consider a goal $\some x\ tuition(x)$.   The system in Section 3  requests the user to select the current 
major.  After the major -- 
 say, med -- is selected,  the system eventually
produces the amount, \ie, $x = 40K$.

\section{An Alternative Operational Semantics}\label{sec:0627}

Our execution model in the previous section, although efficient,  has a serious drawback: it requests the user to perform some actions in advance
even when an execution leads to a failure. 
Fixing this problem requires fundamental changes to our execution model and
adds additional complexity to the model.

To be precise, the new execution model -- adapted from \cite{Jap03} -- now requires two phases:

\begin{numberedlist}

\item the proof phase: This phase builds a proof tree. This proof tree 
 encodes all the possible execution sequences.

\item the execution phase: This phase requests the user to choose
 one execution sequence among all  the  possible execution sequences.

\end{numberedlist}

Given a program $\Delta$ and a goal $G$, a proof tree of $nil;;\Delta \supset G$ is a list of tuples of
the form $\lb E,i \rb$ or $\lb E,(i,j) \rb$ where $E$ is a proof formula ( a formula + execution mode) and $i,j$ are the distances to $F$'s chilren
in the proof tree. Below the execution mode $\pra(\Pscr,\Delta,G)$ is represented by a
proof formula of the form $\Pscr;;\Delta \supset G$, $\prs(D,\Pscr,A)$ by
$D;\Pscr \supset A$ and $\prg(\Pscr,G)$ by $\Pscr\supset G$.
In addition, $a_1::\ldots::a_n::nil$ represents a list of $n$ elements.

\begin{defn}\label{def:semantics}
Let $G$ be a goal and let $\Pscr,\Delta$ be a program that consists of only reusable clauses.
In addition, assume that $\Pscr$ consists of only Horn clauses.
Then the task of proving $\Pscr;;\Delta \supset G$ and returning its proof tree $L$ -- 
written as $\pr(\Pscr;;\Delta \supset G,L)$ -- is defined as follows:

\begin{numberedlist}

\item $\pr(E,\lb E,1 \rb::L)$
if  $\pr(\{ !C \} \cup \Pscr;;\Delta \supset  G,L)$
where $E$ is $\Pscr;;(!C::\Delta)\supset G$ \% process a Horn clause in $D^\Uparrow$ mode

\item    $\pr(E,\lb E,(m+1,1) \rb::L_2)$ if 
 $\pr(\Pscr;;(D_0::\Delta)\supset G,L_0)$  and $\pr(\Pscr;;(D_1::\Delta)\supset  G, L_1)$
and $append(L_0,L_1,L_2)$ and  $length(L_1,m)$. 
 where $E$ is $\Pscr;;((D_0 \add D_1)::\Delta) \supset G$ \% process a choice-disjunctive
clause in  $D^\Uparrow$ mode

\item    $\pr(E,\lb E,1 \rb::L)$ if 
 $\pr(\Pscr\supset G,L)$. where $E$ is $\Pscr;;nil\supset G$
\% switch from   $D^\Uparrow$ mode   to goal processing  mode

\item  $\pr(E,\lb E, - \rb::nil)$
where $E$ is $A;\Pscr\supset A$. \% This is a leaf node in $D^\Downarrow$ mode

\item    $\pr(E, \lb E,1\rb::L)$ if 
 $\pr(\Pscr\supset G_0,L)$ where $E$ is $(G_0\supset A);\Pscr\supset A$.
\%  backchaining in $D^\Downarrow$ mode

\item    $\pr(E, \lb E,1\rb::L) $ if   $\pr([t/x]D;
\Pscr\supset A,L) $ where   $E$ is $\all x D;\Pscr\supset A$. \%  in $D^\Downarrow$ mode

\item    $\pr(E,\lb E,1\rb::L)$ if   $D \in \Pscr$ and $\pr(D;\Pscr\supset A,L)$ where $E$ is $\Pscr\supset A$. \% proving an atomic goal

\item $\pr(E,\lb E,(m+1,1) \rb::L_2 )$ 
 if $\pr(\Pscr\supset G_0,L_0)$ and 
$\pr(\Pscr\supset G_1,L_1)$ and $append(L_0,L_1,L_2)$ and $length(L_1,m)$. 
where $E$ is $\Pscr\supset G_0 \otimes G_1$. \% proving a conjunctive goal

 \item $\pr(E,\lb E,1\rb::L)$  if $\pr(\Pscr\supset [t/x]G ,L)$ where $E$ is $\Pscr\supset \some x G$.

\end{numberedlist}
\end{defn}
\noindent Here, $\Pscr$ is initialized to an empty list.

Once a proof tree is built, the execution phase actually solves the goal relative to the program 
using the proof tree.  Below A $choose$ B means that the machine must choose a successful 
one between the task A and the task B.

\begin{defn}\label{def:exec}
Let $i$ be an index and let $L$ be a proof tree.  
Then   executing $L_i$ (the $i$ element in $L$) -- written as $\prove(i,L)$ --
 is defined as follows: 

\begin{numberedlist}

\item  $\prove(i,L)$ if $L_i = (E,-)$. \% success

\item  $\prove(i,L)$ if $L_i = (\Pscr\supset  G_0 \otimes G_1, (n,m))$ and
 $\prove(i-n,L)$ and $\prove(i-m,L)$. \% the case of two children

\item  $\prove(i,L)$ if $L_i = (\Pscr;;(D_0 \add D_1)::\Delta \supset G, (n,m))$ and
 $read(k)$ and (($k == 0$ and $\prove(i-n,L))$ choose $(k == 1$ and $\prove(i-m,L)))$. \%
In the case of choice disjunctive clauses, the machine requests the user to choose one
between $D_0$ and $D_1$ and then executes the chosen path.

\item  $\prove(i,L)$ if $L_i = (E,1)$  and $\prove(i-1,L)$. \% the remaining cases

\end{numberedlist}
\end{defn}

\noindent  
Now given a program $\Delta$ and a goal $G$, $L$ is initialized to the proof tree of  
$nil;;\Delta \supset G$ and $n$ is initialized to the length of $L$.

\section{Conclusion}\label{sec:conc}

In this paper, we have considered an extension to Prolog with  
\cdc\ in linear logic. This extension allows 
clauses of 
the form  $D_0 \add  D_1$   where $D_0, D_1$ are  Horn clauses.
In particular, these clauses  make it possible for  Prolog
to model decision steps from the user. 

At this stage, clauses of more complex forms such  as $! (D\add D)$ or $(D \land D)$
are not allowed in our language. We plan to allow them in the future.
We also plan to connect our execution model to Japaridze's expressive Computability Logic \cite{Jap03,Jap08}
 in the near future.

%\section{Acknowledgements}

%This work  was supported by Dong-A University Research Fund.

\bibliographystyle{plain}

%\profile*{}{}% without picture of author's face

%\end{multicols}

\end{document}